\documentclass{egpubl}
\usepackage{eurovis2019}

\SpecialIssuePaper         

 \electronicVersion 

\ifpdf \usepackage[pdftex]{graphicx} \pdfcompresslevel=9
\else \usepackage[dvips]{graphicx} \fi

\PrintedOrElectronic

\usepackage{t1enc,dfadobe}

\usepackage{egweblnk}
\usepackage{cite}

\usepackage{amsfonts}
\usepackage{graphicx}
\usepackage[table]{xcolor}
\usepackage{color}
\usepackage{tabularx}
\usepackage{booktabs}
\usepackage{stackengine}
\usepackage{subfig}
\usepackage{tikz}
\usetikzlibrary{arrows}

\usepackage{microtype}
\usepackage{todonotes}
\usepackage{amsmath}
\usepackage[export]{adjustbox}
\usepackage[normalem]{ulem}

\newcommand{\vc}{{\mathbf c}}
\newcommand{\vd}{{\mathbf d}}

\newcommand{\vt}{{\mathbf t}}

\newcommand{\vv}{{\mathbf v}}

\newcommand{\vx}{{\mathbf x}}
\newcommand{\xx}{{\mathbf x}}

\newcommand{\vNull}{{\mathbf 0}}

\newcommand{\mI}{{\mathbf I}}
\newcommand{\mJ}{{\mathbf J}}

\newcommand{\mQ}{{\mathbf Q}}

\newcommand{\Transp}{{{\mathrm T}}}

\newcommand{\cN}{{\mathcal N}}

\newcommand{\diff}{{\mathrm d}}

\newcommand{\dataset}[1]{\textsc{#1}}

\newcommand{\ie}{i.e.,~}
\newcommand{\eg}{e.g.,~}

\newcommand\review[1]{{#1}}

\newcommand\Fig[1]{Fig.~\ref{fig:#1}}
\newcommand\Sec[1]{Section~\ref{sec:#1}}
\newcommand\Eq[1]{Eq.~\eqref{eq:#1}}

\definecolor{bkcolor}{RGB}{210,10,210}
\definecolor{tgcolor}{RGB}{123,50,210}

\title[Robust Reference Frame Extraction from Unsteady 2D Vector Fields with CNNs]%
      {Robust Reference Frame Extraction from Unsteady 2D Vector Fields with Convolutional Neural Networks}

\author[Byungsoo Kim \& Tobias G{\"u}nther]
{\parbox{\textwidth}{\centering Byungsoo Kim 
        and Tobias G{\"u}nther 
        }
        \\
{\parbox{\textwidth}{\centering Department of Computer Science, ETH Zurich }
}
}

\begin{document}

\teaser{
	\begin{tikzpicture}%
	\node[anchor=south west,inner sep=0] (imageI) at (0,1.5) {\includegraphics[width=0.47\linewidth]{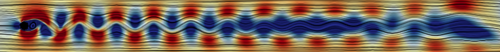}};
	\node[anchor=south west,inner sep=0] (imageG) at (9.3,1.5) {\includegraphics[width=0.47\linewidth]{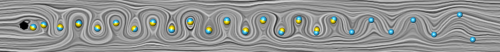}};
	\node[anchor=south west,inner sep=0] (imageN) at (0,0) {\includegraphics[width=0.47\linewidth]{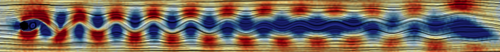}};
	\node[anchor=south west,inner sep=0] (imageC) at (9.3,0) {\includegraphics[width=0.47\linewidth]{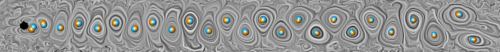}};
	\begin{scope}[x={(imageI.south east)},y={(imageI.north west)}]
	\node[draw=none,overlay,black,anchor=south west] at (0.0,0.98) {\tiny vector field\vphantom{p} (magnitude color-coded)};
	\end{scope}
	\begin{scope}[x={(imageN.south east)},y={(imageN.north west)}]
	\node[draw=none,overlay,black,anchor=south west] at (0.0,0.98) {\tiny vector field + noise and resampling artifacts};
	\end{scope}
	\begin{scope}[x={(imageG.south east)},y={(imageG.north west)}]
	\node[draw=none,overlay,black,anchor=south west] at (0.01,0.98) {\tiny linear optimization \cite{Guenther17:Objective}};
	\end{scope}
	\begin{scope}[x={(imageC.south east)},y={(imageC.north west)}]
	\node[draw=none,overlay,black,anchor=south west] at (0.01,0.98) {\tiny our CNN-based approach};
	\end{scope}
	\path (imageI.south) edge[->, thick, -latex] node [] {} (imageN.north);
	\path (imageN.east) edge[out=0, in=180, ->, thick, -latex] node [] {} (imageG.west);
	\path (imageN.east) edge[out=0, in=180, ->, thick, -latex] node [] {} (imageC.west);
	\end{tikzpicture}
  \centering
   \caption{We developed a novel CNN-based reference frame extraction algorithm that is trained to handle inputs with noise and resampling artifacts. Compared to a linear reference frame optimization~\cite{Guenther17:Objective}, our method is more robust to artifacts. The input vector field (w/ and w/o noise) is shown on the left, and the extraction of vortex centers (orange and yellow), compared to a ground truth (blue) is shown on the right. By combining filtering and reference frame extraction via CNNs, vortex extraction becomes more robust. 
	Note that in the experiment above, the CNN has not seen the cylinder flow during training. In fact, it only trained on a synthetic data base that we introduce in the paper.}
 \label{fig:teaser}
}

\maketitle
\begin{abstract}
Robust feature extraction is an integral part of scientific visualization.
In unsteady vector field analysis, researchers recently directed their attention towards the computation of near-steady reference frames for vortex extraction, which is a numerically challenging \review{endeavor}.
In this paper, we utilize a convolutional neural network to combine two steps of the visualization pipeline in an end-to-end manner: the filtering and the feature extraction.
We use neural networks for the extraction of a steady reference frame for a given unsteady 2D vector field.
By conditioning the neural network to noisy inputs and resampling artifacts, we obtain numerically stabler results than existing optimization-based approaches. 
Supervised deep learning typically requires a large amount of training data.
Thus, our second contribution is the creation of a vector field benchmark data set, which is generally useful for any local deep learning-based feature extraction.
Based on Vatistas velocity profile, we formulate a parametric vector field mixture model that we parameterize based on numerically-computed example vector fields in near-steady reference frames.
Given the parametric model, we can efficiently synthesize thousands of vector fields that serve as input to our deep learning architecture.
The proposed network is evaluated on an unseen numerical fluid flow simulation.
\def\preprintversion{} 
\ifx\preprintversion\undefined

\else

  \small
  \vspace{3mm}
  This is the authors preprint. The definitive version is available at 
	https://onlinelibrary.wiley.com/ and at https://diglib.eg.org/.
  \vspace{2mm}
\fi
\end{abstract}  
\section{Introduction}

The robust extraction of vortices remained to this day one of the most difficult problems of unsteady vector field analysis~\cite{Guenther18:VortexSTAR}.
Vortices themselves are studied in many different applications, such as for engine design~\cite{Roth96,Garth07:Swirl}, blood flow analysis~\cite{Koehler13,Oeltze16} or even in the atmosphere of other planets~\cite{Hadjighasem16}.
A vortex is commonly understood as a set of particles rotating around a common point or axis, if the flow is viewed in the correct reference frame~\cite{Lugt79,Robinson91}.
This dependence on the reference frame is what makes vortex extraction in unsteady flows difficult.
On the one hand, existing methods concentrated on vortex characterizations that give the same result for many different reference frames, such as those arising from a Galilean transformation~\cite{Okubo70,Hunt87,Jeong95,Weinkauf07} or from a smooth rotation and translation of the reference frame~\cite{Astarita79,Haller05,Haller16LAVD}.
On the other hand, reference frames have been searched in which topologically relevant structures appear~\cite{Bhatia14b}.
Lugt~\cite{Lugt79} and Robinson~\cite{Robinson91} argued that the relevant reference frame is the one in which the flow becomes steady.
Thus, recently G{\"u}nther et al.~\cite{Guenther17:Objective} formulated the local search for the steady reference frame as an optimization problem.
Their method, however, relies on a regularization that often prevents them from finding a perfectly steady solution.
In the limit case of no regularization, their solution depends on second-order derivatives that are extremely difficult to estimate robustly if the data is noisy.
Robustness to noise and resampling artifacts is also a major concern when it comes to the local vortex extraction algorithms~\cite{Globus91,Peikert99} that are applied in the resulting frame.
An algorithm for the robust extraction of the optimal reference frame that is insensitive to noise and other distortion artifacts is desperately needed in order to achieve a reliable vortex detection in unsteady vector fields.

Inspired from the recent success of machine learning approaches in robustly solving image processing tasks~\cite{Goodfellow16:DeepLearning}, we propose to train a convolutional neural network (CNN) to locally extract the optimal reference frame, in which a given unsteady 2D vector field becomes steady.
We thereby concentrate on the robustness of the extractor by training it on synthetically distorted data, which includes the adding of noise and a prior resampling.
A general challenge of machine learning is that it typically requires a large amount of data to successfully generalize.
Since machine learning is a rather young field for flow visualization, we are not aware of any large benchmark vector fields that could directly be used for training.
A second contribution of our paper is therefore the synthetic generation and the release of a vector field benchmark data set that may be used in the future by other researchers, looking for applications of machine learning in flow visualization.
In summary:
\begin{itemize}
	\item We introduce a novel parametric flow mixture model that is based on Vatistas~\cite{Vatistas91} velocity profiles. After fitting parameters to numerical data, we synthesize a large vector field collection that contains divergence-free and compressible flows.
	\item We design and evaluate a convolutional neural network to locally extract the optimal reference frame in which a given 2D vector field becomes steady.
\end{itemize}
In terms of performance our method is slightly slower than a linear optimization~\cite{Guenther17:Objective}, but our CNN-based approach has up to 3-4 times lower reconstruction residuals under increasing noise and resampling artifacts, making our method much more robust on real-world data.
Fig.~\ref{fig:teaser} gives a first impression of the feature extraction results that are possible on distorted data after computing the optimal reference frame with either our approach or the linear optimization~\cite{Guenther17:Objective}.

\section{Related Work}\label{sec:relatedwork}

In the following sections, we lay the foundations for our work, including a summary of visualization methods that utilize machine learning, and an overview of objective vortex extraction methods.

\subsection{Machine Learning in Visualization}
\paragraph*{Explainable AI.}
Machine learning and visualization have great potential for synergies.
Explainable AI is currently a highly relevant research area that uses visualization to look into the black box that deep learning is often considered to be.
We refer to Seifert et al.~\cite{Seifert17:DeepLearningSurveyVision}, Hohman et al.~\cite{Hohman18:VisualAnalyticsDeepLearning} and Ancona et al.~\cite{Ancona18:Attribution} for an introduction into this promising area.

\paragraph*{Deep Learning for Visualization.}
The opposite direction, \ie the use of deep learning to solve visualization tasks is still fairly uncharted territory.
Frey~\cite{Frey17} trained a neural network to determine the best progressive sampling strategy to calculate the similarity of spatio-temporal data sets. 
In information visualization, Fan and Hauser~\cite{Fan18} used CNNs to improve the manual brushing of points in scatterplots. 
Berger et al.~\cite{Berger17} explored the use of generative adversary networks to analyze the role of transfer functions in the image synthesis process of direct volume rendering. 
For a steady 3D vector field, Han et al.~\cite{Han18} used an autoencoder to learn a low-dimensional feature space of a voxel representation of randomly generated streamlines or stream surfaces.
They plotted the low-dimensional feature space with t-SNE and selected representative geometric primitives from a density-based clustering. 

\paragraph*{CNNs for Vortex Extraction.}
Lguensat et al.~\cite{Lguensat17} used a classification CNN to identify ocean eddies from sea surface height maps.
Not only interested in the detection of ocean eddies but also in their tracking, Franz et al.~\cite{Franz18:OceanEddyCNN} trained a classification CNN that receives the Okubo-Weiss~\cite{Okubo70,Weiss91} criterion as input and tracked the structures with optical flow and a spatio-temporal recurrent neural network.
Bin and Li~\cite{Bin18} classified normalized vector field patches of size $9\times 9$ with a classification CNN into clockwise rotating, counterclockwise rotating, saddle type and others.
Str{\"o}fer et al.~\cite{Strofer18} trained a classification CNN to identify specific fluid flow features in the domain, such as recirculation regions, boundary layers and a horseshoe vortex.
Deng et al.~\cite{Deng18:IVD-CNN} trained a CNN to recall the instantaneous vorticity deviation (IVD) of Haller et al.~\cite{Haller16LAVD}. 

All methods above either performed a classification or produced a region-based vortex measure.
Since we have line-based vortex extraction in mind, numerical stability is of greater concern to us.
Thus, we utilize CNNs to greatly improve the numerical stability by conditioning the network to noisy and resampled inputs, resulting is the first deep learning-based reference frame extraction.

\subsection{Objective Vortex Extraction}
Over the past decades, dozens of vortex extraction algorithms have been proposed in the flow visualization and fluid mechanics literature.
We refer to G{\"u}nther and Theisel for a recent overview~\cite{Guenther18:VortexSTAR}.

\subsubsection{Definition of Objectivity}
The most recent vortex extraction methods~\cite{Haller16LAVD,Guenther17:Objective,Guenther18:InertialObjective,Guenther19:Affine,Hadwiger19} aspired to be \emph{objective}.
A measure is called \emph{objective} if it remains invariant under a smooth rotation and/or smooth translation of the reference frame~\cite{Truesdell65}.
Such a transformation transforms a point $(\vx,t)$ to the location $(\vx^\ast,t^\ast)$ with:
\begin{align}
\vx^\ast = \mQ(t)\vx + \vc(t)  \;,~\quad  t^\ast = t - a  \label{eq:deepvort-trafo-point}
\end{align}
for a time-dependent rotation matrix $\mQ(t)$, a time-dependent translation vector $\vc(t)$ and a constant time shift $a$.
Objectivity guarantees that a measure looks the same for different rotations and translations of the observer.
Because of the relativity between the motion of the vortex and the motion of the observer, this is equivalent to the guarantee that the measure of a moving feature always look the same if the feature rotates and/or translates.
Objectivity is therefore useful, when searching for moving features in unsteady flows.

\subsubsection{Reference Frame Transformation of a Vector Field}\label{sec:reference-transform}
In order to train a network that recovers a reference frame transformation $\mQ(t)$, $\vc(t)$ in which a vector field appears steady, we have to generate a large number of unsteady vector fields with a known ground truth transformation.
For this, we start from a steady vector field and transform it with many random reference frame transformations, which makes each resulting flow unsteady.
In general, a vector field $\vv(\xx,t)$ is transformed into a new frame via~\cite{Guenther17:Objective}:
\begin{align}
\vv^\ast(\xx^\ast,t^\ast) = \mQ(t)\vv(\xx,t) + \frac{\diff \mQ(t)}{\diff t}\xx + \frac{\diff \vc(t)}{\diff t}	\label{eq:deepvort-trafo-velocity}
\end{align}
The evaluation of this equation requires the transformation of $(\xx^\ast,t^\ast)$ into the old frame $(\xx,t)$ in order to sample the given (steady) vector field $\vv(\xx,t)$ at the correct location.
Since $\mQ(t)$ is orthogonal, rearranging~\Eq{deepvort-trafo-point} gives:
\begin{align}
\xx = \mQ(t)^\Transp \left(\xx^\ast - \vc(t)\right)  \;,~\quad  t = t^\ast + a  \label{eq:deepvort-trafo-point-inverse}
\end{align}
By inserting~\Eq{deepvort-trafo-point-inverse} into~\Eq{deepvort-trafo-velocity}, we can uniformly sample the space-time domain $(\xx^\ast,t^\ast)$ of the transformed unsteady vector field onto a regular grid, which is later fed with the corresponding transformation $\mQ(t)$ and $\vc(t)$ to the neural network during training.

\subsubsection{Objective Region-based Vortex Extraction Methods}
In order to derive an objective version of the vorticity tensor, Drouot and Lucius~\cite{Drouot76} and later Tabor and Klapper~\cite{Tabor94} independently proposed the \emph{relative vorticity tensor}, which is viewed in the strain rate basis.
The strain rate tensor is objective and thus, all measures viewed in its eigenvector basis become objective as well.
Astarita~\cite{Astarita79} formally proved the objectivity of this approach and further introduced an objective index measure that classifies the domain into regions that either perform extension-like motions or rigid-body-like rotations.
Haller~\cite{Haller05} introduced the $M_z$ criterion, which first uses the strain rate acceleration tensor to classify the domain into elliptic and hyperbolic regions.
By taking a Lagrangian perspective, their approach identifies coherent vortices using particles that stay for a long time in elliptic areas, \ie they perform rotating motions.
Vortex boundaries are also seen as elliptic Lagrangian coherent structures~\cite{Haller2015:LCS}. 
More recently, Haller et al.~\cite{Haller16LAVD} identified coherent vortices objectively using the original vorticity tensor $\Omega$.
They introduced a Eulerian measure called instantaneous vorticity deviation (IVD), which subtracts the spatial mean vorticity of the neighborhood, and the integral of IVD along a pathline called Lagrangian averaged vorticity deviation.

\subsubsection{Reference Frame Optimization}
An objective vortex coreline extractor has recently been proposed by G{\"u}nther et al.~\cite{Guenther17:Objective}.
Since we use their method as baseline in our comparison, we briefly explain the approach in more detail.
In the late 1970s, Lugt~\cite{Lugt79} characterized vortices as closed or spiraling streamlines in a reference frame in which the flow field becomes steady.
Similarly, Robinson~\cite{Robinson91} identified them as closed streamlines in a reference frame that moves with the vortex center.
While it was clear early on that not a single reference frame exists in which all vortices become steady~\cite{Lugt79}, since they might move into different directions, it was recognized that the frame has to be searched locally~\cite{Perry94}.
For this reason, G{\"u}nther et al.~\cite{Guenther17:Objective} solved an optimization problem to find a local reference frame in which the vector field becomes steady.
Their goal was to minimize the time partial of the transformed field $\vv^\ast$ earlier shown in~\Eq{deepvort-trafo-velocity} in a local neighborhood $U$:
\begin{align}
\int_{U} \left\| \frac{\partial \vv^\ast(\xx^\ast, t)}{\partial t} \right\|^2 dV  \rightarrow min
\label{eq:time_partial}
\end{align}
Thereby, the unknowns are the rotation $\mQ(t)$ and the translation $\vc(t)$ of the reference frame.
Important to us is how the rotation $\mQ(t)$ and the translation $\vc(t)$ can be discretized.
When computing the time partial of $\vv^\ast$ in~\Eq{deepvort-trafo-velocity} we only need up to second-order partials~\cite{Guenther17:Objective}.
Further, we can restrict the transformation at time $t$ to $\mQ(t)=\mI$ and $\vc(t)=\vNull$ to obtain our vortex features at the same locations as in the input field $\vv$.
Thus, the only unknowns are the first-order and second-order derivatives of the rotation and translation, evaluated at time $t$:
\begin{align}
\dot\mQ = \frac{\diff \mQ(t)}{\diff t}  \;,\;
\ddot\mQ = \frac{\diff^2 \mQ(t)}{\diff t^2}  \;,\;
\dot\vc = \frac{\diff \vc(t)}{\diff t}  \;,\;
\ddot\vc = \frac{\diff^2 \vc(t)}{\diff t^2}
\nonumber
\end{align}
In 2D, these are six numbers (two angles and two 2D vectors).
G{\"u}nther et al.~\cite{Guenther17:Objective} have shown that this minimization can be efficiently solved as a linear optimization problem.
Recently, they extended their method to affine transformations~\cite{Guenther19:Affine}.
Meanwhile, Hadwiger et al.~\cite{Hadwiger19} formulated the search as a global optimization problem. 
They avoided the choice of a suitable neighborhood and instead regularized the method by the assumption of isometry, \ie the reference frame transformation is given by an approximate Killing field.
\review{
Prior to the search for steady reference frames, other approaches have been followed to find a distinguished reference frame for unsteady flow, including the subtraction of the harmonic vector field component found by a Helmholtz-Hodge decomposition~\cite{Wiebel04,Wiebel07:LocalizedFlow,Bhatia14b} and reference frames derived from a Galilean-invariant topology that is based on an analysis of the determinant of the Jacobian~\cite{bujack2016topology}.
}

\section{Synthetic Generation of Vector Fields}
\label{sec:parametric-model}
We propose an \review{end-to-end approach that combines preprocessing (smoothing) and feature extraction (reference frame extraction)} to compute for \review{a possibly noisy} unsteady vector field the reference frame transformation in which the flow becomes steady.
To train the model, we first synthetically generate a large data base, containing thousands of vector field patches.

\subsection{Parametric Mixture Model for Vector Fields}
If enough diverse real-world training data was available, variational autoencoders~\cite{Kingma13:VAE} or generative adversarial networks~\cite{Goodfellow14:GAN} could be used to synthesize further training data.
In our work, we incorporate experimental observations~\cite{Vatistas91} to formulate an explicit parametric vector field mixture model that is specialized on vortices.
As shown later, it generalizes to other flow features.
Our analytic model is readily available to everyone, and works without a large flow data base.

\subsubsection{Vatistas Vortex Velocity Profile}

\begin{figure}[b]%
\includegraphics[width=\columnwidth]{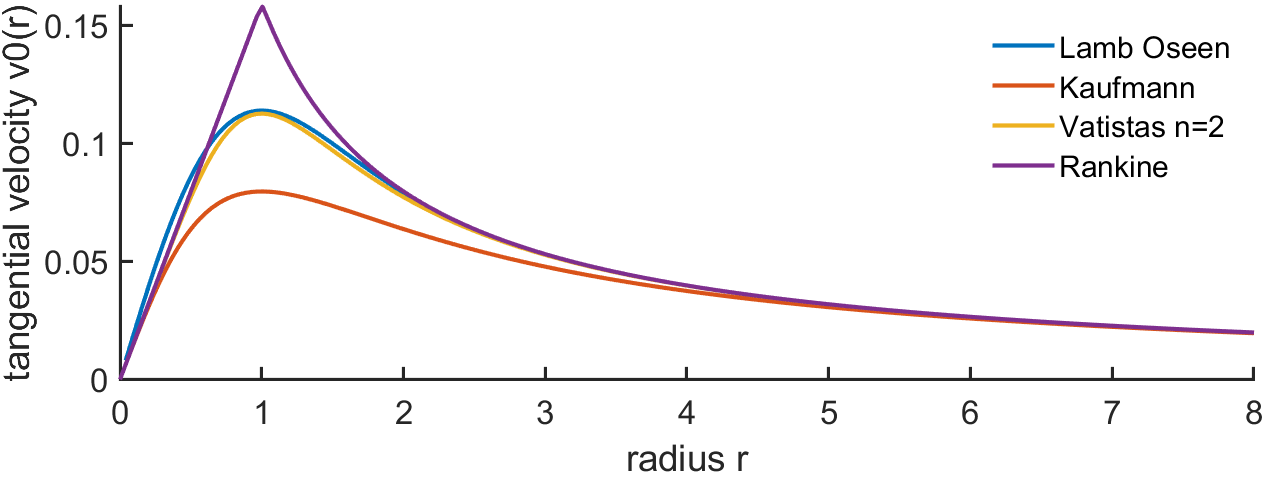}%
\vspace{-2mm}%
\caption{Plot of common vortex velocity profiles. Starting from the center, the tangential velocity increases up until a certain maximum (here at $r_c=1$). Afterwards, it decays with increasing distance. The Vatistas model~\cite{Vatistas91} contains Kaufmann~\cite{Kaufmann62} for $n=1$ and Rankine for $n\rightarrow\infty$ as special cases. For $n=2$, Vatistas is similar to the Lamb-Oseen model~\cite{Oseen1912}.}%
\label{fig:vortex-models}%
\end{figure}

Since we are later mainly interested in vortices, we make use of Vatistas~\cite{Vatistas91} experimentally-obtained vortex velocity profile. 
With $r_c$ being the radius with maximal velocity, the tangential flow velocity $v_0(r)$ of a rotationally-symmetric unit vortex is:
\begin{align}
v_0 (r) = \frac{r}{ 2\pi r_c^2 \left( (\frac{r}{r_c})^{2n} + 1\right)^{\frac{1}{n}} } ~. 
\label{eq:vatistas}
\end{align}
This equation was taken from Bhagwat and Leishman~\cite{Bhagwat00}, who formulated the model for varying vortex radii.
Note that Vatistas' vortex model contains other well-known vortex models as special cases, such as the Kaufmann vortex~\cite{Kaufmann62} (for $n=1$) and the Rankine vortex (for $n\rightarrow\infty$).
Thus for increasing $n$, the behavior near the critical point becomes linear.
For $n=2$, Vatistas model is similar to the Lamb-Oseen vortex model~\cite{Oseen1912}.
Note that Eq.~\eqref{eq:vatistas} denotes the radius $r$ in absolute values, rather than being relative to $r_c$ as in the literature~\cite{Bhagwat00}.
The different vortex models are shown in Fig.~\ref{fig:vortex-models}.
We leave the core radius $r_c$ and the shape exponent $n$ as degrees of freedom in the parametric model.

\subsubsection{Parametric Mixture Model}
\label{sec:deep-mixture-model}
\begin{figure}%
\begin{minipage}[t]{\columnwidth}
\centering%
$|\vv_p|:$~~ 0 \includegraphics[width=0.4\columnwidth]{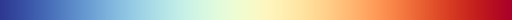} 0.15 ~~~~~~~~~~
\end{minipage}\vspace{1mm}\\
\begin{minipage}[t]{0.315\columnwidth}
\includegraphics[width=\columnwidth]{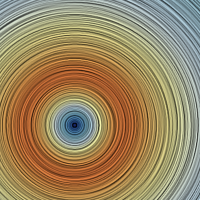}\\%
\small$\vc$\,=\,$(1,1)$, $\vd$\,=\,$(0,0)$, $\vt$\,=\,$($-$\frac{1}{2},$-$\frac{1}{2})$, $r_c$\,=\,$1$, $n$\,=\,$2$
\end{minipage}\hfill%
\begin{minipage}[t]{0.315\columnwidth}
\includegraphics[width=\columnwidth]{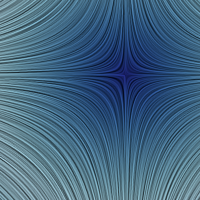}\\%
\small$\vc$\,=\,$(0,0)$, $\vd$\,=\,$(1,-1)$, $\vt$\,=\,$(\frac{1}{2},\frac{1}{2})$, $r_c$\,=\,$3$, $n$\,=\,$8$
\end{minipage}\hfill%
\begin{minipage}[t]{0.315\columnwidth}
\includegraphics[width=\columnwidth]{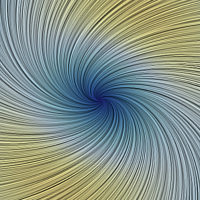}\\%
\small$\vc$\,=\,$(1,\frac{1}{2})$, $\vd$\,=\,$(1,1)$, $\vt$\,=\,$(0,0)$, $r_c$\,=\,$2$, $n$\,=\,$2$
\end{minipage}%
\caption{Examples of steady 2D flows, generated by our model in Eq.~\eqref{eq:def-vp}. Here, shown for the spatial domain $\mathcal{D} = [-2,2]^2$ .}%
\label{fig:vortex-model-instances}%
\end{figure}

Based on Vatistas velocity profile $v_0(r)$ in Eq.~\eqref{eq:vatistas}, we define a steady flow primitive $\vv_p$ 
with a critical point at $\vt=(t_x,t_y)$ as:
\begin{align}
\vv_p(x,y) = 
\begin{bmatrix}
	d_x & c_x \\ -c_y & d_y 
\end{bmatrix} \begin{pmatrix}
	x - t_x \\ y - t_y
\end{pmatrix} \cdot \frac{v_0(\sqrt{(x-t_x)^2+(y-t_y)^2})}{\sqrt{(x-t_x)^2+(y-t_y)^2}}
\label{eq:def-vp}
\end{align}
with the physical meaning that $\vc=(c_x,c_y)$ describes the vortical motion and $\vd=(d_x,d_y)$ denotes the in-flow and out-flow.
In total, each primitive has eight degrees of freedom: $c_x$, $c_y$, $d_x$, $d_y$, $t_x$, $t_y$, $r_c$, and $n$.
Depending on the parameterization, a range of different flow structures appear, including vortices, sinks, sources and saddles.
Different instances of the model are shown in Fig.~\ref{fig:vortex-model-instances}.
By restricting the above parameters, it would be possible to incorporate additional domain knowledge, for instance, by concentrating on divergence-free flows only.
We used the full range of all eight parameters to test the limits of a general model.

Since our vector field patches might contain multiple critical points, we use a mixture model of $m$ flow primitives:
\begin{align}
\vv(x,y) = \sum_{p=1}^m \vv_p(x,y) \;,
\label{eq:def-vmixture}
\end{align}
which in general results in $8\,m$ parameters.
To find physically plausible parameter configurations, we fit the above mixture model to patches of a numerical data set, which we describe next.
\review{Other approaches to formulate a model are discussed later in Section~\ref{sec:discussion}.}

\subsubsection{Parameter Space Fitting}
Eq.~\eqref{eq:def-vmixture} gives rise to a large number of different vector fields, which leads to two questions to answer: how can we restrict the parameter space to physically-plausible flows and are there flows that cannot be captured by the model?
To answer these questions, we fit the above model to numerically simulated vector fields.
Since we are aiming for a vector field collection that represents flows in optimal near-steady reference frames, we first extract the optimal reference frame with the approach of G{\"u}nther et al.~\cite{Guenther17:Objective}.
Since the parameters in Eq.~\eqref{eq:def-vmixture} are non-linear, we first use 200 iterations of simulated annealing, which is followed by further 200 iterations of gradient descent to settle into the local minimum. 
Fitting a vector field requires a distance metric between vector fields.
In this paper, we use an $L_1$ distance with gradients, which was introduced by Kim et al.~\cite{kim18b} in the context of velocity reconstruction using CNNs. 
They introduced two metrics: one for divergence-free flows and one for compressible flows.
Since we do not constrain the parameter range in our parametric model in Eq.~\eqref{eq:def-vp}, we opt for the more general distance metric for compressible flows:
\begin{align}
\mathcal{L}(\hat{\vv}, \vv) = ||\hat{\vv}- \vv||_1+\lambda||\nabla\hat{\vv}- \nabla\vv||_1,
\label{eq:gradient-loss}
\end{align}
where $\hat{\vv}$ and $\vv$ are the vector fields to compare, and $\lambda$ is a weight for the gradient difference. 
In practice, we use $\lambda=1$, as it gives lower residuals than for $\lambda=0$ or for the $L_2$-norm (Mean Squared Error), as shown in Appendix~\ref{sec:parameterfitting}.
Fig.~\ref{fig:fitting-cylinder-heatmap} shows a heat map in the \dataset{Cylinder} flow, displaying for each voxel in the domain, how closely we could fit the numerical data with our parameteric mixture model.
It becomes apparent that obstacles are not well represented in the model, but the remainder of the domain is approximated well.
Below in Fig.~\ref{fig:fitting-cylinder-images}, individual patches of the domain are shown, displaying how closely the mixture model matches the numerical data for the varying numbers of model components $m$.
For $m=3$, we achieve a high accuracy, and we thus use $m=3$ models later during the vector field synthesis.
Finally, Fig.~\ref{fig:fitting-cylinder-histogram} shows histograms of the individual parameters.
The histograms are used in the next section to sample further vector fields.
In Appendix~\ref{sec:app-boussinesq}, additional fitting results are shown for the \dataset{Boussinesq} vector field.

\begin{figure}[t]%
\subfloat[Heat maps of the fitting residual for varying number of mixture model components $m$. The higher $m$, the lower the residual. Obstacles are not yet contained in the parametric flow model.\label{fig:fitting-cylinder-heatmap}]{
\begin{minipage}[t]{\linewidth}
\centering
$|1-\hat{\vv} / \vv|:$~~ 0 \includegraphics[width=0.4\columnwidth]{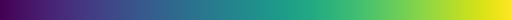} 1 ~~~~~~~~~~
\raisebox{0.12in}{\rotatebox[origin=t]{90}{$m=1$}}~%
\includegraphics[width=0.95\columnwidth]{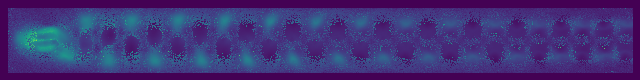}\vspace{0.5mm}\\%
\raisebox{0.12in}{\rotatebox[origin=t]{90}{$m=2$}}~%
\includegraphics[width=0.95\columnwidth]{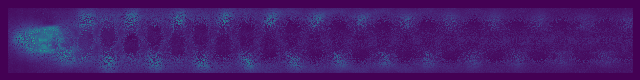}\vspace{0.5mm}\\%
\raisebox{0.12in}{\rotatebox[origin=t]{90}{$m=3$}}~%
\includegraphics[width=0.95\columnwidth]{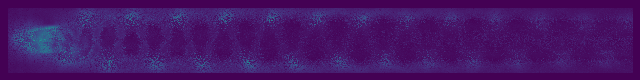}%
\end{minipage}}\\%
\subfloat[Fitting results for selected vector field patches. The fitting improves with more degrees of freedom, i.e., for a higher number of models $m$.\label{fig:fitting-cylinder-images}]{
\begin{minipage}[t]{\linewidth}
\includegraphics[width=0.24\columnwidth]{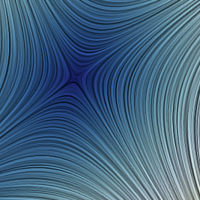}\hfill%
\includegraphics[width=0.24\columnwidth]{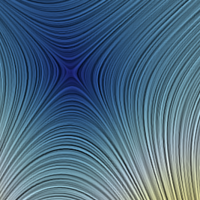}\hfill%
\includegraphics[width=0.24\columnwidth]{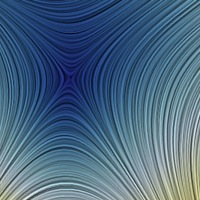}\hfill\hfill%
\includegraphics[width=0.24\columnwidth]{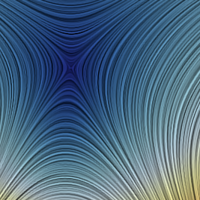}\vspace{0.5mm}\\%
\includegraphics[width=0.24\columnwidth]{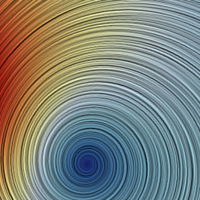}\hfill%
\includegraphics[width=0.24\columnwidth]{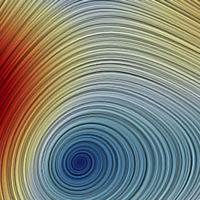}\hfill%
\includegraphics[width=0.24\columnwidth]{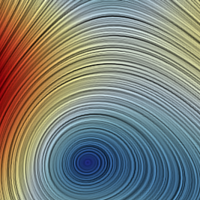}\hfill\hfill%
\includegraphics[width=0.24\columnwidth]{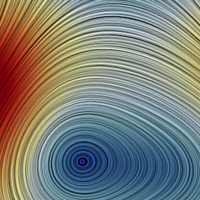}\vspace{-1mm}\\%
\begin{minipage}[t]{0.24\columnwidth} \centering \small $m=1$ \end{minipage}\hfill%
\begin{minipage}[t]{0.24\columnwidth} \centering \small $m=2$ \end{minipage}\hfill%
\begin{minipage}[t]{0.24\columnwidth} \centering \small $m=3$ \end{minipage}\hfill\hfill%
\begin{minipage}[t]{0.24\columnwidth} \centering \small reference \end{minipage}%
\end{minipage}}\\%
\subfloat[Histograms of the individual model parameters, showing the near-Gaussian distribution of the individual parameter values.\label{fig:fitting-cylinder-histogram}]{
\includegraphics[width=\columnwidth]{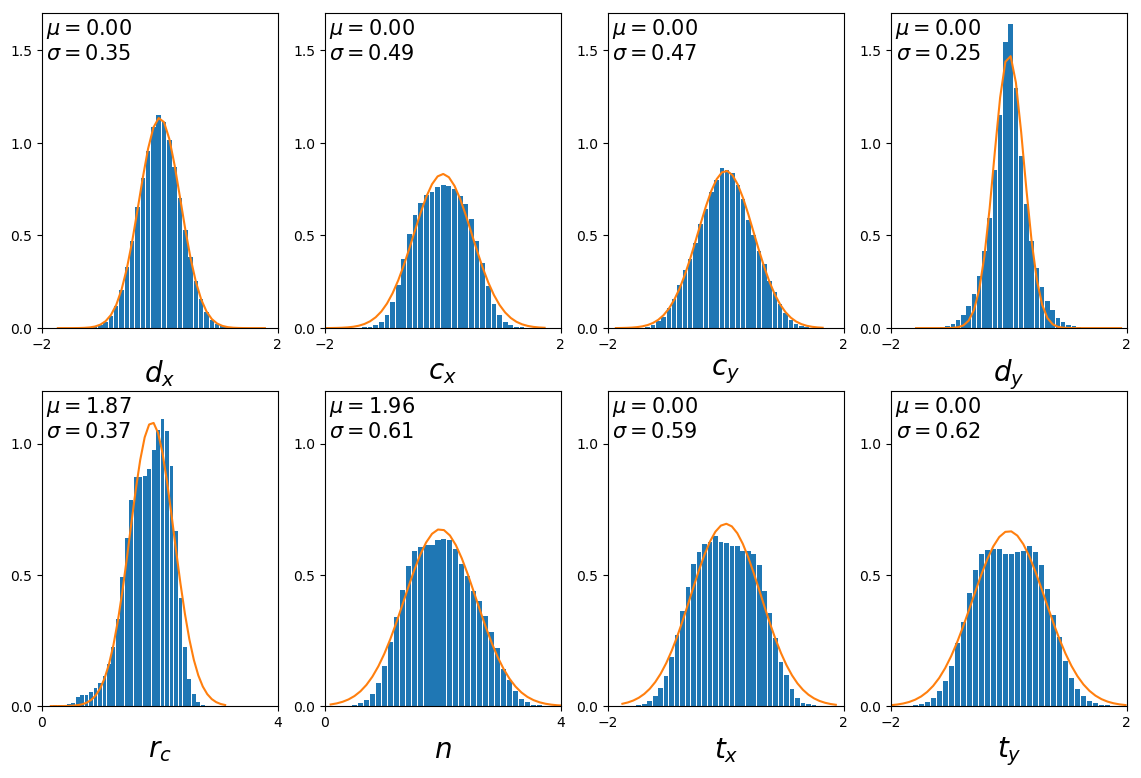}%
}
\caption{Fitting results for the \dataset{Cylinder} flow, showing that our parametric flow mixture model approximates numerical data well.}%
\label{fig:fitting-cylinder}%
\end{figure}

\subsubsection{Sampling of the Parameter Space}
\label{sec:sampling-space}
The histograms in Fig.~\ref{fig:fitting-cylinder-histogram} characterize the distribution of model parameters needed in order to generate further vector field patches of similar type.
We individually approximate the distribution of each model parameter ($c_x$, $c_y$, $d_x$, $d_y$, $t_x$, $t_y$, $r_c$, and $n$) by a separate Gaussian distribution $\cN$. 
The mean and standard deviation of each component are shown in Fig.~\ref{fig:fitting-cylinder-histogram}.
In order to sample more flows, similar to the \dataset{Cylinder} flow, the correlation between the random variables would have to be accounted for.
In order to span a wider range of vector fields, we purposefully neglect the correlations and sample each variable independently.
This means, we sample each Gaussian in Fig.~\ref{fig:fitting-cylinder-histogram}, and insert the resulting parameters into the parametric model in Eq.~\eqref{eq:def-vmixture}, resulting in a new analytic vector field.
Since the sampling process is computationally cheap, we can quickly \review{synthesize} thousands of vector field patches \review{in order to train a convolutional neural network for reference frame extraction}, as we describe next.

\section{Deep Learning of Reference Frame Extraction}
\label{sec:CNN-reference-frames}

\subsection{Overview}
In this paper, we propose a convolutional neural network that spans two stages of the visualization pipeline: the filtering and the feature extraction.
By combining both \review{in an end-to-end} fashion, the \review{reference frame} extractor can easily be conditioned to handle noisy inputs and data with resampling artifacts.
During our supervised learning, we teach the network pairs of unsteady vector fields and their corresponding optimal reference frame transformation. 
These training examples are generated by transforming a steady vector field into an unsteady reference frame.
By distorting the input data, the network learns to undo artifacts, which greatly improves robustness and helps in subsequent feature extraction tasks, such as the detection of vortex centers, which we demonstrate later.
The following sections introduce the network architecture and the generation of the training data in more detail.

\subsection{Architecture}
\begin{figure}[h]
\centering
\includegraphics[width=0.48\textwidth]{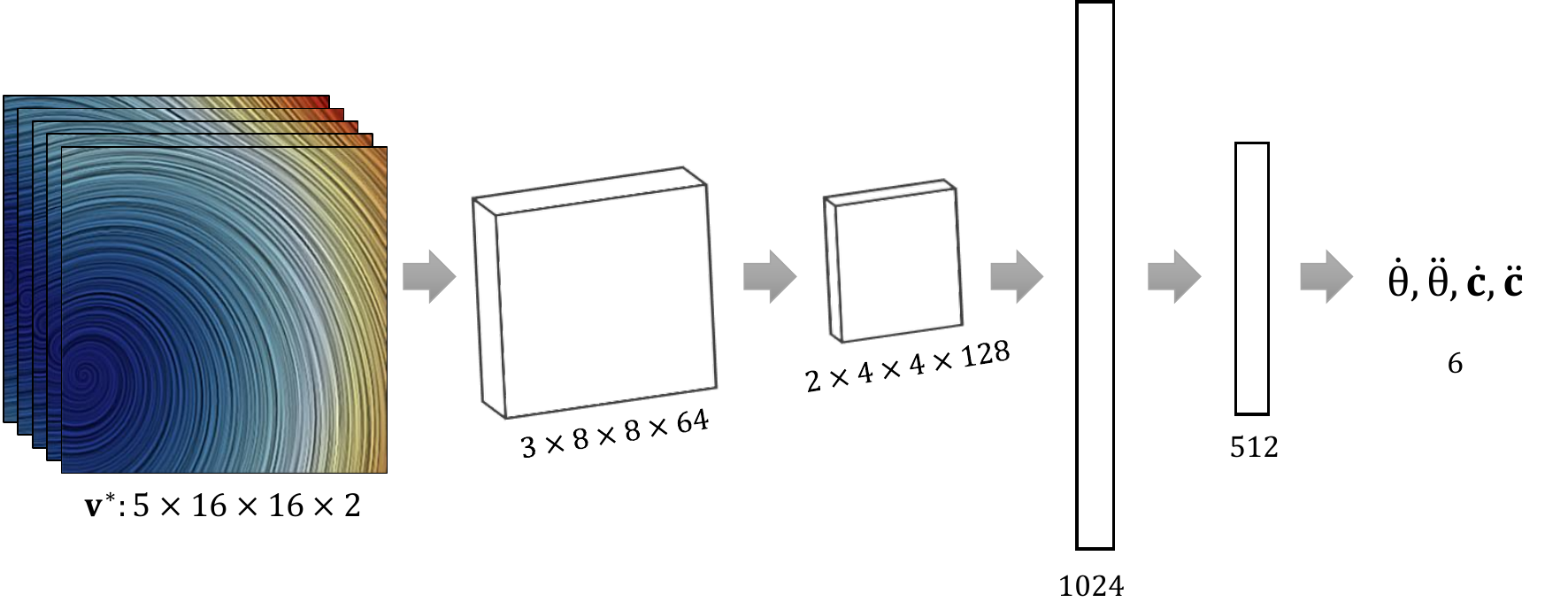}
\caption{Our CNN Architecture. The numbers below each box represent the dimension of feature maps.}
\label{fig:architecture}
\end{figure}

We build our network as a form of typical CNN as seen in \Fig{architecture}. The first part of our CNN consists of 3D convolutional kernels. In this part, the dimension reduction is performed for the feature extraction with a kernel size of $3\times3\times3$ and strides of $2\times2\times2$, followed by a batch normalization and a rectified linear unit (ReLU) layer. The size of the filter is doubled starting from 64 to maximally 1024. With input 2D unsteady fields $\vv^\ast \in \mathbb{R}^{T \times H \times W \times 2}$, the number of convolutional layers is calculated by $n = log_2{\max(H,W)}-2$, and the output dimension of last convolutional layer would be $max(\frac{T}{2^n},1) \times \frac{H}{2^n} \times \frac{W}{2^n} \times min(64^n, 1024)$.
The second part of our network exploits fully connected layers instead of convolutional filters for the final inference from identified high-level vortex features. Batch normalization and a ReLU layer are followed same as convolutional layers, and a dropout with the probability of 0.1 is used to avoid overfitting. 
Note that we infer the first and second-order derivatives of the reference frame transformation $\dot\mQ, \ddot\mQ, \dot\vc, \ddot\vc$ in 2D. Here we denote them as a 6-dimensional parameter vector $[\dot \theta, \ddot \theta, \dot x, \dot y, \ddot x, \ddot y]$.

\subsection{Data Sets}\label{sec:datasets}
We synthesize our data set based on Vatistas velocity profile as described in Section~\ref{sec:parametric-model}. Once a 2D steady field is generated by a super-position of three models with sampled parameters, we build unsteady vector fields by applying a reference frame transformation, followed by degrading with \review{varying degrees of additive} uniform noise \review{(i.e., white noise)} 
and down-up resampling.
\review{For simplicity we chose white noise during training, but we also test the network on a data set that was corrupted with Gaussian noise.}

\subsubsection{Training Data}
\begin{figure}
\begin{tikzpicture}%
\node[anchor=north west,inner sep=0] (11s) at (0,5.8)     {\includegraphics[width=0.3\linewidth]{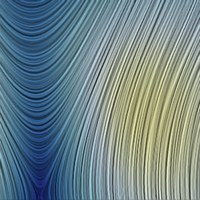}};
\node[anchor=north west,inner sep=0] (12u5) at (2.85,6.2) {\includegraphics[width=0.3\linewidth,frame]{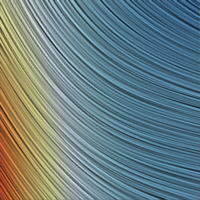}};
\node[anchor=north west,inner sep=0] (12u4) at (2.80,6.1) {\includegraphics[width=0.3\linewidth,frame]{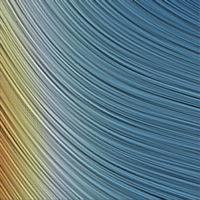}};
\node[anchor=north west,inner sep=0] (12u3) at (2.75,6)   {\includegraphics[width=0.3\linewidth,frame]{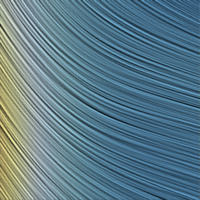}};
\node[anchor=north west,inner sep=0] (12u2) at (2.70,5.9) {\includegraphics[width=0.3\linewidth,frame]{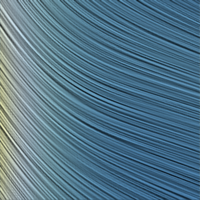}};
\node[anchor=north west,inner sep=0] (12u1) at (2.65,5.8) {\includegraphics[width=0.3\linewidth,frame]{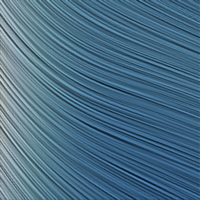}};
\node[anchor=north west,inner sep=0] (13u5) at (5.70,6.2) {\includegraphics[width=0.3\linewidth,frame]{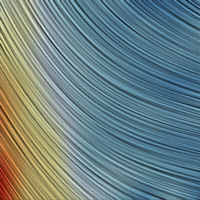}};
\node[anchor=north west,inner sep=0] (13u4) at (5.65,6.1) {\includegraphics[width=0.3\linewidth,frame]{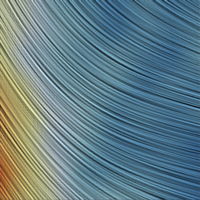}};
\node[anchor=north west,inner sep=0] (13u3) at (5.60,6)   {\includegraphics[width=0.3\linewidth,frame]{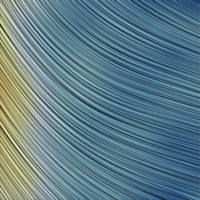}};
\node[anchor=north west,inner sep=0] (13u2) at (5.55,5.9) {\includegraphics[width=0.3\linewidth,frame]{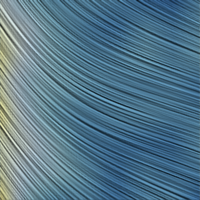}};
\node[anchor=north west,inner sep=0] (13u1) at (5.50,5.8) {\includegraphics[width=0.3\linewidth,frame]{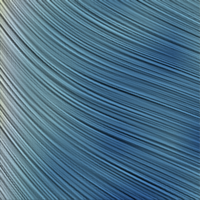}};

\node[anchor=north west,inner sep=0] (21s) at (0,2.8)     {\includegraphics[width=0.3\linewidth]{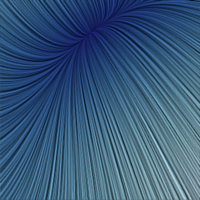}};
\node[anchor=north west,inner sep=0] (22u5) at (2.85,3.2) {\includegraphics[width=0.3\linewidth,frame]{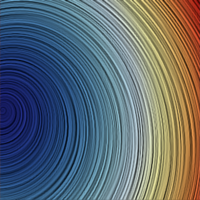}};
\node[anchor=north west,inner sep=0] (22u4) at (2.80,3.1) {\includegraphics[width=0.3\linewidth,frame]{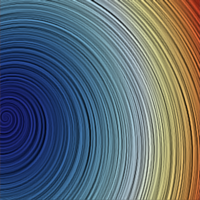}};
\node[anchor=north west,inner sep=0] (22u3) at (2.75,3)   {\includegraphics[width=0.3\linewidth,frame]{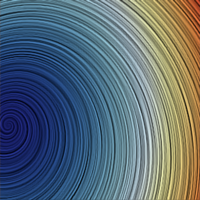}};
\node[anchor=north west,inner sep=0] (22u2) at (2.70,2.9) {\includegraphics[width=0.3\linewidth,frame]{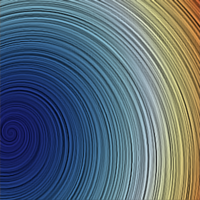}};
\node[anchor=north west,inner sep=0] (22u1) at (2.65,2.8) {\includegraphics[width=0.3\linewidth,frame]{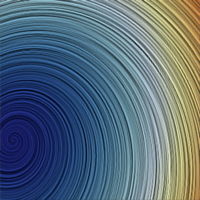}};
\node[anchor=north west,inner sep=0] (23u5) at (5.70,3.2) {\includegraphics[width=0.3\linewidth,frame]{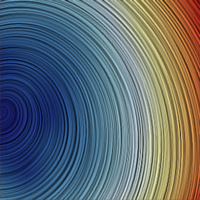}};
\node[anchor=north west,inner sep=0] (23u4) at (5.65,3.1) {\includegraphics[width=0.3\linewidth,frame]{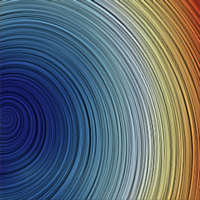}};
\node[anchor=north west,inner sep=0] (23u3) at (5.60,3)   {\includegraphics[width=0.3\linewidth,frame]{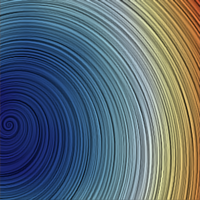}};
\node[anchor=north west,inner sep=0] (23u2) at (5.55,2.9) {\includegraphics[width=0.3\linewidth,frame]{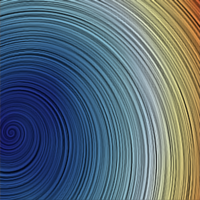}};
\node[anchor=north west,inner sep=0] (23u1) at (5.50,2.8) {\includegraphics[width=0.3\linewidth,frame]{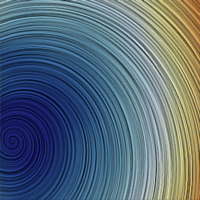}};

\node[anchor=north west,inner sep=0] (31s) at (0,-0.2)       {\stackunder{\includegraphics[width=0.3\linewidth]{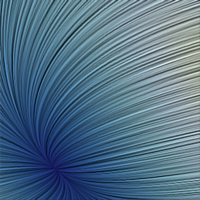}}{\footnotesize{Steady Field}}};
\node[anchor=north west,inner sep=0] (32u5) at (2.85,0.2) {\includegraphics[width=0.3\linewidth,frame]{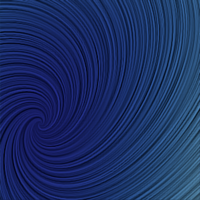}};
\node[anchor=north west,inner sep=0] (32u4) at (2.80,0.1) {\includegraphics[width=0.3\linewidth,frame]{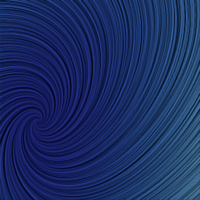}};
\node[anchor=north west,inner sep=0] (32u3) at (2.75,0)   {\includegraphics[width=0.3\linewidth,frame]{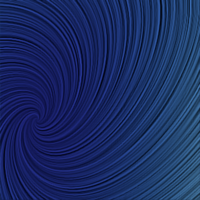}};
\node[anchor=north west,inner sep=0] (32u2) at (2.70,-0.1){\includegraphics[width=0.3\linewidth,frame]{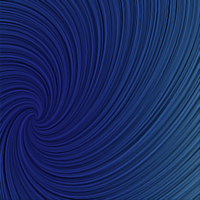}};
\node[anchor=north west,inner sep=0] (32u1) at (2.65,-0.2){\stackunder{\includegraphics[width=0.3\linewidth,frame]{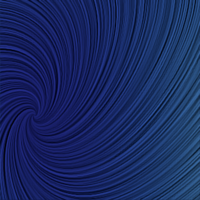}}{\footnotesize{Unsteady Field}}};
\node[anchor=north west,inner sep=0] (33u5) at (5.70,0.2) {\includegraphics[width=0.3\linewidth,frame]{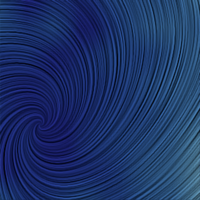}};
\node[anchor=north west,inner sep=0] (33u4) at (5.65,0.1) {\includegraphics[width=0.3\linewidth,frame]{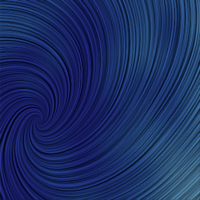}};
\node[anchor=north west,inner sep=0] (33u3) at (5.60,0)   {\includegraphics[width=0.3\linewidth,frame]{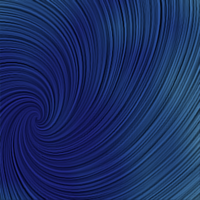}};
\node[anchor=north west,inner sep=0] (33u2) at (5.55,-0.1){\includegraphics[width=0.3\linewidth,frame]{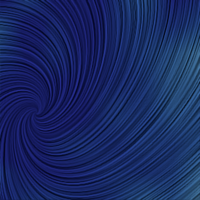}};
\node[anchor=north west,inner sep=0] (33u1) at (5.50,-0.2){\stackunder{\includegraphics[width=0.3\linewidth,frame]{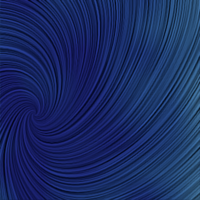}}{\footnotesize{Degenerated Field}}};
\end{tikzpicture}
\centering
\caption{Examples of training data pairs. From left to right, steady field, transformed unsteady field and distorted field. You can see some spots on magnitude plots of degenerated flows.}
\label{fig:datasets}
\end{figure}

As described in Section~\ref{sec:reference-transform}, we transform each 2D steady vector field to an unsteady 2D vector field.
Note that parameters for the rotation and translation $[\dot \theta, \ddot \theta, \dot x, \dot y, \ddot x, \ddot y]$ are uniformly sampled.
Then, we degrade it by adding uniform noise and down-up resampling, which is designed to simulate a general degeneration in a data acquisition system.
The number of samples is 30,000 and the data set for training and testing is split into the ratio of 9:1. 
Some examples of the training data are illustrated in Fig.~\ref{fig:datasets}.
We refer to \Sec{parameters} for details about each operation.

\subsubsection{Numerical Data for Validation}
After training, we have validated our neural network on the \dataset{Cylinder} data set. As the input window size of our CNN is fixed, we have evaluated this data set by sliding a lookup window over the entire domain. In order to evaluate the \dataset{Cylinder} flow, which has a resolution of $5\times80\times640$, we slide a lookup window of size $5\times16\times16$ with the stride 1 and sequentially predict parameters for the center pixel. (\eg 159 steps for 40,625 windows with a batch size of 256).

\paragraph*{Patch Normalization}
Since the training data set is defined in the 2D unit domain (\ie $\mathcal{\bar X} \times \mathcal{\bar Y} \times \mathcal{\bar T} = [-1,1]^3$), input patches of numerical data sets are also scaled into this domain.
Thus, given velocity patch slices $\vv_p : \mathcal{X} \times \mathcal{Y} \times \mathcal{T} \rightarrow \mathcal{X} \times \mathcal{Y}$, defined in the domain $\mathcal{X} \times \mathcal{Y} \times \mathcal{T} = [x_{min},x_{max}] \times [y_{min},y_{max}] \times [t_{min},t_{max}]$, we compute $\bar\vv_p$ in the unit domain via:
\begin{align}
\bar{\vv}_p(\xx) = 
\begin{bmatrix}
	\frac{t_{max}-t_{min}}{x_{max}-x_{min}} & 0 \\ 0 & \frac{t_{max}-t_{min}}{y_{max}-y_{min}} 
\end{bmatrix}
\vv_p(\xx).
\end{align}
Furthermore, we globally normalize the magnitude of the training data to $[-1,1]$ during training. Thus, re-scaled patches from the validation data set are finally normalized by the same factors that have been applied to the training data before it was fed to the network.

\subsection{Implementation}
We implemented our convolutional neural network (CNN) using Keras~\cite{chollet2015keras} with Tensorflow~\cite{abadi2016tensorflow} as backend. The networks are trained on each data set for 300 epochs using an Adam optimizer~\cite{kingma2014adam} with a learning rate of 0.001 and the mean squared error (MSE) loss function. We chose a batch size of 256. Finally, the model is selected, which shows the minimum test error during training.
All visualizations were created with the visualization toolkit Amira~\cite{Stalling05}.

\section{Result}\label{sec:result}
In the following sections, we first apply our network to synthetic data generated with our parametric mixture model, before testing it on the reference frame extraction from unseen numerical data. Afterwards, the performance is discussed.

\subsection{Training and Testing on Synthetic Data}
\begin{figure}[t]
\begin{tabular}{*{3}{c@{\hspace{3px}}}}
\includegraphics[width=0.14\textwidth]{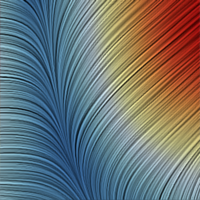} &
\includegraphics[width=0.14\textwidth]{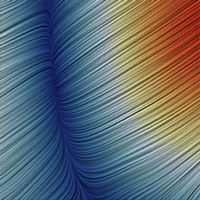} &
\includegraphics[width=0.14\textwidth]{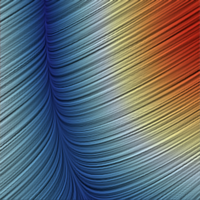} \\

\includegraphics[width=0.14\textwidth]{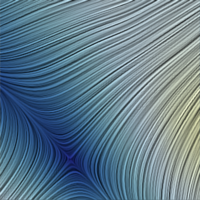} &
\includegraphics[width=0.14\textwidth]{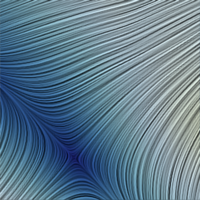} &
\includegraphics[width=0.14\textwidth]{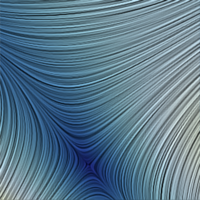} \\

\stackunder{\includegraphics[width=0.14\textwidth]{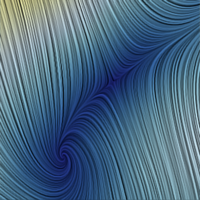}}{\footnotesize{Linear opt. (1.09e-2)}} &
\stackunder{\includegraphics[width=0.14\textwidth]{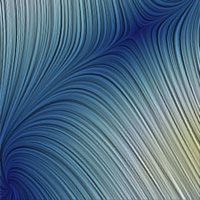}}{\footnotesize{Ours (2.13e-4)}}  &
\stackunder{\includegraphics[width=0.14\textwidth]{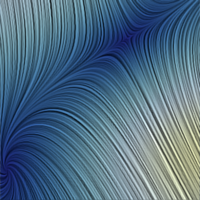}}{\footnotesize{Ground Truth}} \\
\end{tabular}
\caption{Results of vortex extraction on test splits. The numbers in parenthesis \review{are} MSE, and our method robustly handles degenerated flows where linear optimization method fails.}
\label{fig:result}
\end{figure}

As baseline, we compare our method with the linear reference frame optimization by G\"unther et al.~\cite{Guenther17:Objective} on a synthetic data set, generated as described in Section~\ref{sec:sampling-space} with a $1\%$ noise-to-signal ratio in the normalized domain.
We compute the mean-squared error on the inferred parameter vector of the test split, and our CNN shows not only visually more plausible results it also obtains up to two orders of magnitude lower time partial residuals than the linear optimization method, as shown in~\Fig{result}. 
Our method therefore passes the first test: outperform the baseline on synthetic flows similar to the ones seen during training.

\subsection{Validation of Network on Numerical Data}
Next, we apply our network to unseen numerical data to evaluate how well the network generalized, given its synthetic training data.

\subsubsection{Robustness}
By feeding a patch of voxels into the network, the CNN has enough information to learn proper smoothing and filtering kernels, such that noise and resampling artifacts can be compensated.
To test the numerical robustness, we introduce varying degrees of noise and resampling artifacts into the unseen validation data, and compare the network output to the linear optimization.

\paragraph*{Error Metric.} 
As error metric, we use the reference frame transformation 
 obtained by the network or by the linear optimization, respectively, in order to transform the non-distorted unsteady vector field back into its steady frame.
In the resulting frame, we calculate the time partial derivative, which ideally becomes zero, i.e.,
\begin{align}
\vv_t(\xx,t) = \vv_t^\ast(\xx,t) + \dot\mQ\vv(\xx,t) + \ddot\mQ\xx + \ddot\vc - [\mJ(\xx,t)+\dot\mQ]\cdot[\dot\mQ\xx+\dot\vc] 
\end{align}
If $|\vv_t| = 0$, the network successfully found the reference frame transformation, despite the presence of noise.

\begin{figure}%
\subfloat[Plots of the time partial for comparison on different types of flow degeneration. The left plot shows the robustness of each method on different levels of resampling without noise, and the right plot shows the one on different noise levels without resampling.\label{fig:plot-cylinder-images}]{
\begin{minipage}[t]{\linewidth}
\includegraphics[width=0.48\columnwidth]{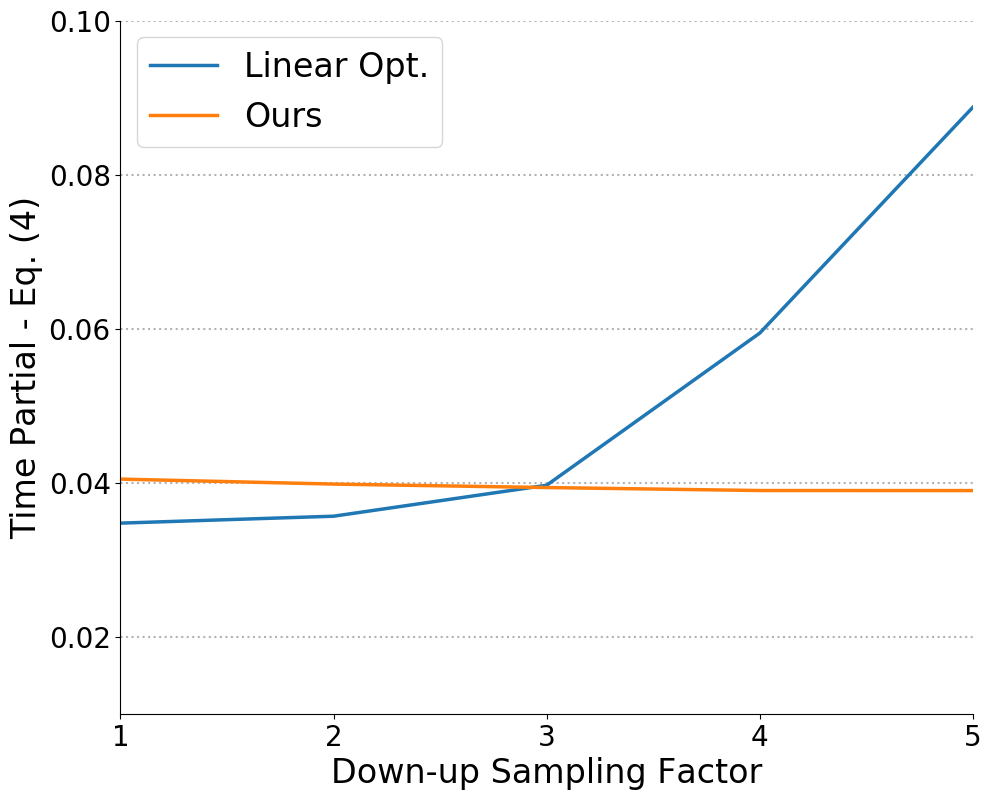}
\includegraphics[width=0.48\columnwidth]{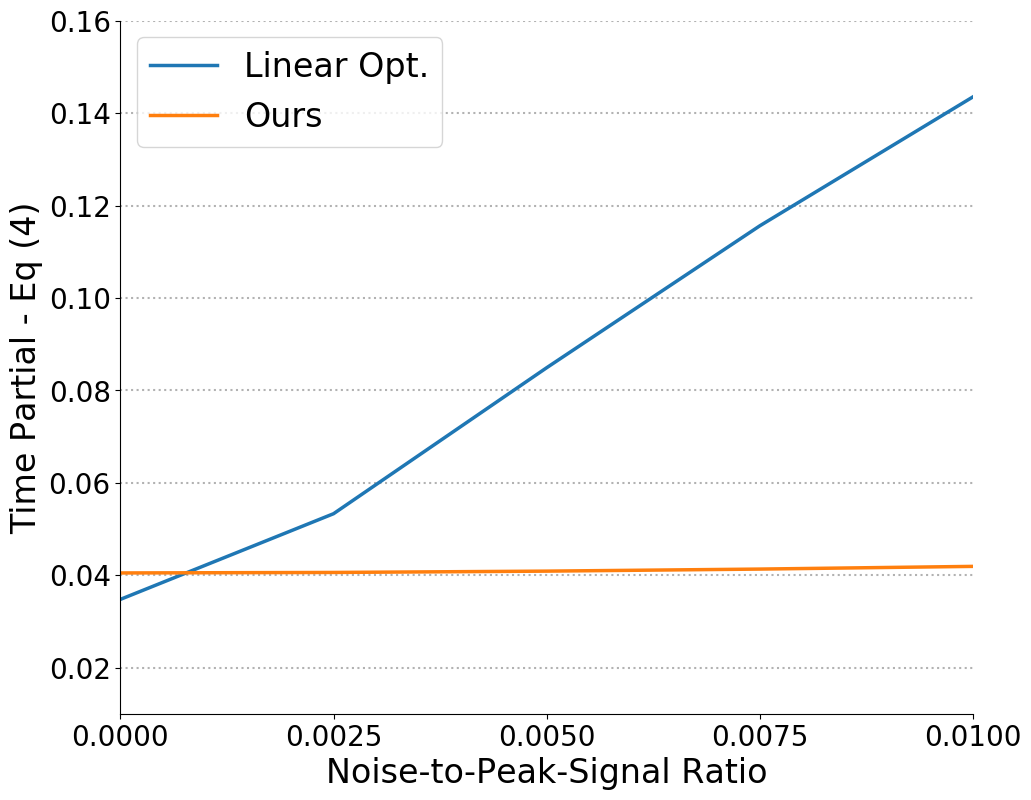}
\end{minipage}}\\\vspace{-2mm}%
\subfloat[Results of the reference frame optimization on the resampled \dataset{Cylinder} data set with the factor of 5. \label{fig:vortex-cylinder-compare-resample}]{
\begin{minipage}[t]{\linewidth}
\centering\vspace{-2mm}%
\raisebox{0.12in}{\rotatebox[origin=t]{90}{\scriptsize Linear Opt.}}~%
\includegraphics[width=0.95\columnwidth]{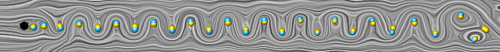}\vspace{0.5mm}\\%
\raisebox{0.12in}{\rotatebox[origin=t]{90}{\scriptsize Ours}}~%
\includegraphics[width=0.95\columnwidth]{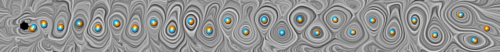}\vspace{0.5mm}%
\end{minipage}}\\%
\subfloat[Results of the reference frame optimization on noise-added \dataset{Cylinder} data set. The ratio of noise to the maximum magnitude of flow is 0.01 ($1\%$). \label{fig:vortex-cylinder-compare-noise}]{
\begin{minipage}[t]{\linewidth}
\centering
\raisebox{0.12in}{\rotatebox[origin=t]{90}{\scriptsize Linear Opt.}}~%
\includegraphics[width=0.95\columnwidth]{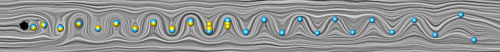}\vspace{0.5mm}\\%
\raisebox{0.12in}{\rotatebox[origin=t]{90}{\scriptsize Ours}}~%
\includegraphics[width=0.95\columnwidth]{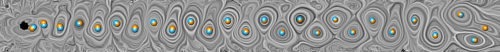}\vspace{0.5mm}\\%
\end{minipage}}\vspace{-1mm}%
\caption{Study of robustness for different levels and types of degeneration in the \dataset{Cylinder} flow. Our method shows robust performances for various degradations. \review{In this experiment, the network was trained for $1\%$ noise and with resampling artifacts.} }%
\label{fig:comparing-cylinder}%
\end{figure}

\paragraph*{Quantitative Experiment.}
We use the \dataset{Cylinder} flow as benchmark.
In this vector field, the swirling strength of the vortices reduces over time, due to numerical dissipation and viscosity, which means that vortices become weaker.
As the angular velocity magnitude decreases down the flow, the influence of the artificially added noise becomes stronger, causing errors in the linear optimization~\cite{Guenther17:Objective} \review{that are discussed later in Section~\ref{sec:discussion} and can be} seen in Fig.~\ref{fig:teaser}.
In consequence, not all vortices are detected.
Since our CNNs have seen resampling artifacts and noise during training, they robustly recover the reference frame in all areas of the domain even for different levels of degeneration as seen in Fig.~\ref{fig:comparing-cylinder}.

\begin{figure}[t]%
\begin{minipage}[t]{\linewidth}
\centering
\raisebox{0.12in}{\rotatebox[origin=t]{90}{\scriptsize uniform}}~%
\includegraphics[width=0.95\columnwidth]{image/robust/vis-network-n}\vspace{0.5mm}\\%
\raisebox{0.12in}{\rotatebox[origin=t]{90}{\scriptsize Gaussian}}~%
\includegraphics[width=0.95\columnwidth]{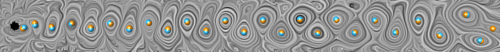}\vspace{0.5mm}%
\end{minipage}\vspace{-2mm}%
\caption{\review{Results of our CNN-based reference frame optimization on two different types of noise in the \dataset{Cylinder} flow. During training only uniform noise was seen by the network ($1\%$ magnitude).}}%
\label{fig:vortex-cylinder-compare-noisetype}%
\end{figure}

\paragraph*{Noise Type Experiment.}
\review{In Fig.}~\ref{fig:vortex-cylinder-compare-noisetype}, \review{we applied our network (trained on $1\%$ uniform noise and with resampling artifacts) to a data set with $1\%$ of Gaussian noise.
Since the results are very similar, our network also seems to handle this type of noise well.
For higher noise magnitudes, differences will start to appear and then it will be better to train the network with Gaussian noise.
}

\begin{figure}[t]%
\subfloat[\review{Results of the reference frame optimization on noise-added \dataset{Cylinder} data set. The ratio of noise to the maximum magnitude of flow is 0.01 (1\%).}\label{fig:vortex-cylinder-compare-noise}]{
\begin{minipage}[t]{\linewidth}
\centering
\raisebox{-0.12in}{\rotatebox[origin=t]{90}{\scriptsize Linear Opt.}}~%
\begin{minipage}[t]{\linewidth}%
\begin{tikzpicture}
		\node[anchor=south west,inner sep=0] (image) at (0,0) {\includegraphics[width=0.95\columnwidth]{image/robust/vis-generic-n}};
		\begin{scope}[node distance=-1.8mm and -1.2mm, x={(image.south east)},y={(image.north west)}]
		\node[draw=none,overlay,white] at (0.06,0.85) {\scriptsize unfiltered};
		\end{scope}
\end{tikzpicture}\vspace{0.5mm}\\%
\begin{tikzpicture}
		\node[anchor=south west,inner sep=0] (image) at (0,0) {\includegraphics[width=0.95\columnwidth]{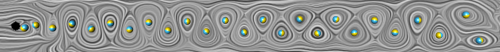}};
		\begin{scope}[node distance=-1.8mm and -1.2mm, x={(image.south east)},y={(image.north west)}]
		\node[draw=none,overlay,white] at (0.09,0.85) {\scriptsize with smoothing};
		\end{scope}
\end{tikzpicture}\vspace{0.5mm}%
\end{minipage}\\%
\raisebox{0.12in}{\rotatebox[origin=t]{90}{\scriptsize Ours \vphantom{p}}}~%
\begin{minipage}[t]{\linewidth}
\begin{tikzpicture}
		\node[anchor=south west,inner sep=0] (image) at (0,0) {\includegraphics[width=0.95\columnwidth]{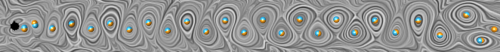}\vspace{0.5mm}};
		\begin{scope}[node distance=-1.8mm and -1.2mm, x={(image.south east)},y={(image.north west)}]
		\node[draw=none,overlay,white] at (0.06,0.85) {\scriptsize unfiltered};
		\end{scope}
\end{tikzpicture}\vspace{0.5mm}
\end{minipage}%
\end{minipage}}\vspace{-1mm}\\
\subfloat[\review{Results of the reference frame optimization on noise-added \dataset{Cylinder} data set. The ratio of noise to the maximum magnitude of flow is 0.05 (5\%).}\label{fig:vortex-cylinder-compare-noise}]{
\begin{minipage}[t]{\linewidth}
\centering
\raisebox{-0.12in}{\rotatebox[origin=t]{90}{\scriptsize Linear Opt.}}~%
\begin{minipage}[t]{\linewidth}%
\begin{tikzpicture}
		\node[anchor=south west,inner sep=0] (image) at (0,0) {\includegraphics[width=0.95\columnwidth]{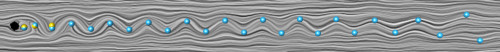}};
		\begin{scope}[node distance=-1.8mm and -1.2mm, x={(image.south east)},y={(image.north west)}]
		\node[draw=none,overlay,white] at (0.06,0.85) {\scriptsize unfiltered};
		\end{scope}
\end{tikzpicture}\vspace{0.5mm}\\%
\begin{tikzpicture}
		\node[anchor=south west,inner sep=0] (image) at (0,0) {\includegraphics[width=0.95\columnwidth]{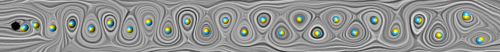}};
		\begin{scope}[node distance=-1.8mm and -1.2mm, x={(image.south east)},y={(image.north west)}]
		\node[draw=none,overlay,white] at (0.09,0.85) {\scriptsize with smoothing};
		\end{scope}
\end{tikzpicture}\vspace{0.5mm}%
\end{minipage}\\%
\raisebox{0.12in}{\rotatebox[origin=t]{90}{\scriptsize Ours \vphantom{p}}}~%
\begin{minipage}[t]{\linewidth}
\begin{tikzpicture}
		\node[anchor=south west,inner sep=0] (image) at (0,0) {\includegraphics[width=0.95\columnwidth]{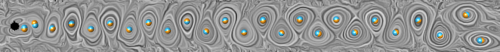}\vspace{0.5mm}};
		\begin{scope}[node distance=-1.8mm and -1.2mm, x={(image.south east)},y={(image.north west)}]
		\node[draw=none,overlay,white] at (0.06,0.85) {\scriptsize unfiltered};
		\end{scope}
\end{tikzpicture}\vspace{0.5mm}
\end{minipage}%
\end{minipage}}\vspace{-1mm}\\
\subfloat[\review{Results of the reference frame optimization on noise-added \dataset{Cylinder} data set. The ratio of noise to the maximum magnitude of flow is 0.1 (10\%).}\label{fig:vortex-cylinder-compare-noise}]{
\begin{minipage}[t]{\linewidth}
\centering
\raisebox{-0.12in}{\rotatebox[origin=t]{90}{\scriptsize Linear Opt.}}~%
\begin{minipage}[t]{\linewidth}%
\begin{tikzpicture}
		\node[anchor=south west,inner sep=0] (image) at (0,0) {\includegraphics[width=0.95\columnwidth]{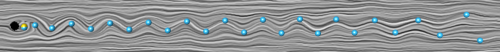}};
		\begin{scope}[node distance=-1.8mm and -1.2mm, x={(image.south east)},y={(image.north west)}]
		\node[draw=none,overlay,white] at (0.06,0.85) {\scriptsize unfiltered};
		\end{scope}
\end{tikzpicture}\vspace{0.5mm}\\%
\begin{tikzpicture}
		\node[anchor=south west,inner sep=0] (image) at (0,0) {\includegraphics[width=0.95\columnwidth]{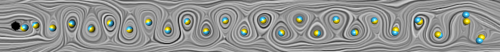}};
		\begin{scope}[node distance=-1.8mm and -1.2mm, x={(image.south east)},y={(image.north west)}]
		\node[draw=none,overlay,white] at (0.09,0.85) {\scriptsize with smoothing};
		\end{scope}
\end{tikzpicture}\vspace{0.5mm}%
\end{minipage}\\%
\raisebox{0.12in}{\rotatebox[origin=t]{90}{\scriptsize Ours \vphantom{p}}}~%
\begin{minipage}[t]{\linewidth}
\begin{tikzpicture}
		\node[anchor=south west,inner sep=0] (image) at (0,0) {\includegraphics[width=0.95\columnwidth]{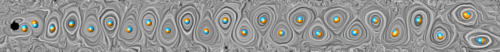}\vspace{0.5mm}};
		\begin{scope}[node distance=-1.8mm and -1.2mm, x={(image.south east)},y={(image.north west)}]
		\node[draw=none,overlay,white] at (0.06,0.85) {\scriptsize unfiltered};
		\end{scope}
\end{tikzpicture}\vspace{0.5mm}
\end{minipage}%
\end{minipage}}\vspace{-1mm}%
\caption{Comparison of robustness of vortex extraction methods on different levels and types of degeneration on \dataset{Cylinder} flow. Our method shows robust performances for various degradation, although it has not seen them during training. \review{For the noise experiments, we also show results of the linear method~\cite{Guenther17:Objective} after applying Gaussian smoothing to the noisy input vector field.}}%
\label{fig:comparing-cylinder-noiselevel}%
\end{figure}

\paragraph*{Noise Magnitude Experiment.}
\review{In order to study the behavior of our network under a stronger influence of noise, we retrained the network and introduced for each patch up to $10\%$ of white noise (uniformly sampled from $0\%$ to $10\%$) and applied our CNN at testing time to data with $1\%$, $5\%$ and $10\%$ of noise.
The results are shown in Fig.~\ref{fig:comparing-cylinder-noiselevel}.
In this comparison, we also tested how much prior Gaussian smoothing (3 iterations using a $7\times 7$ filter kernel) helps the linear method, i.e., the preprocessing is done explicitly before the reference frame extraction.
For small noise magnitudes, smoothing helps, however, smoothing also affects the position of the vortex core. 
Especially for large noise ratios, the CNN-based method recovers the reference frame better.
Without prior smoothing the linear method removes almost no ambient motion, keeping the dominant downstream ambient motion of the input flow.
}

\subsubsection{Temporal Coherence}
\begin{figure*}%
\subfloat[Linear optimization on noisy data.]{
\includegraphics[width=0.3\textwidth]{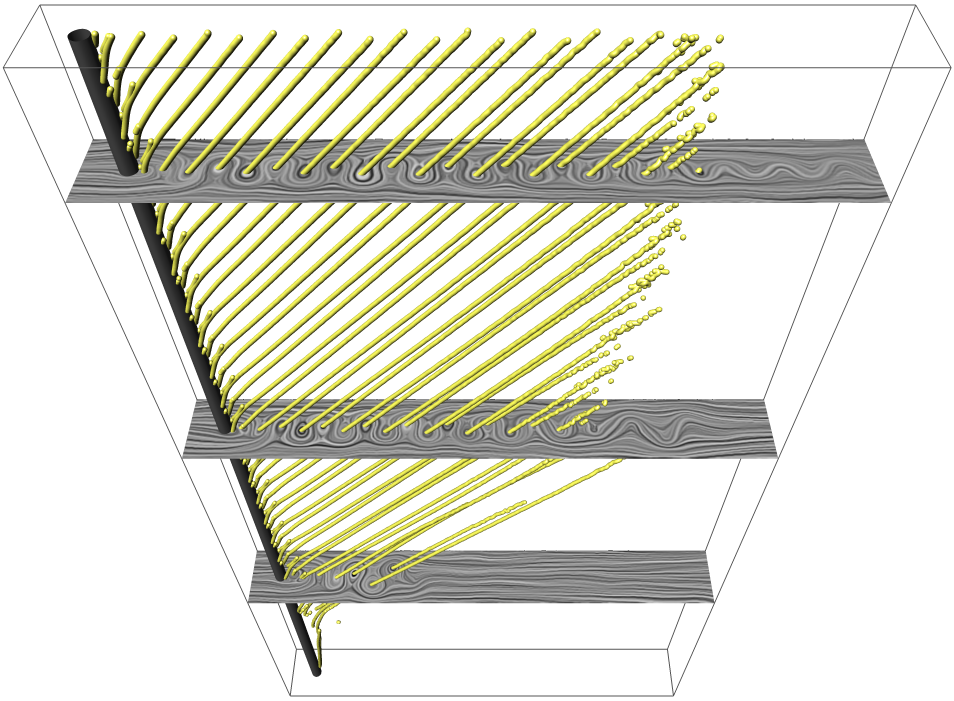}}\hfill%
\subfloat[Our CNN-based approach on noisy data.]{
\includegraphics[width=0.3\textwidth]{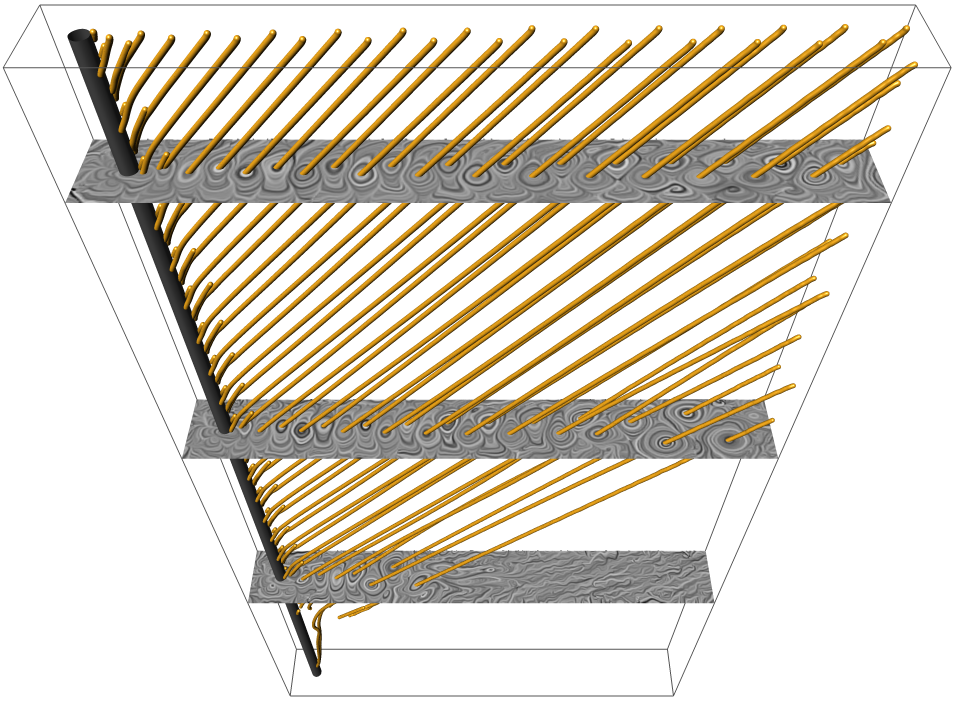}}\hfill%
\subfloat[Linear optimization on original data.]{
\includegraphics[width=0.3\textwidth]{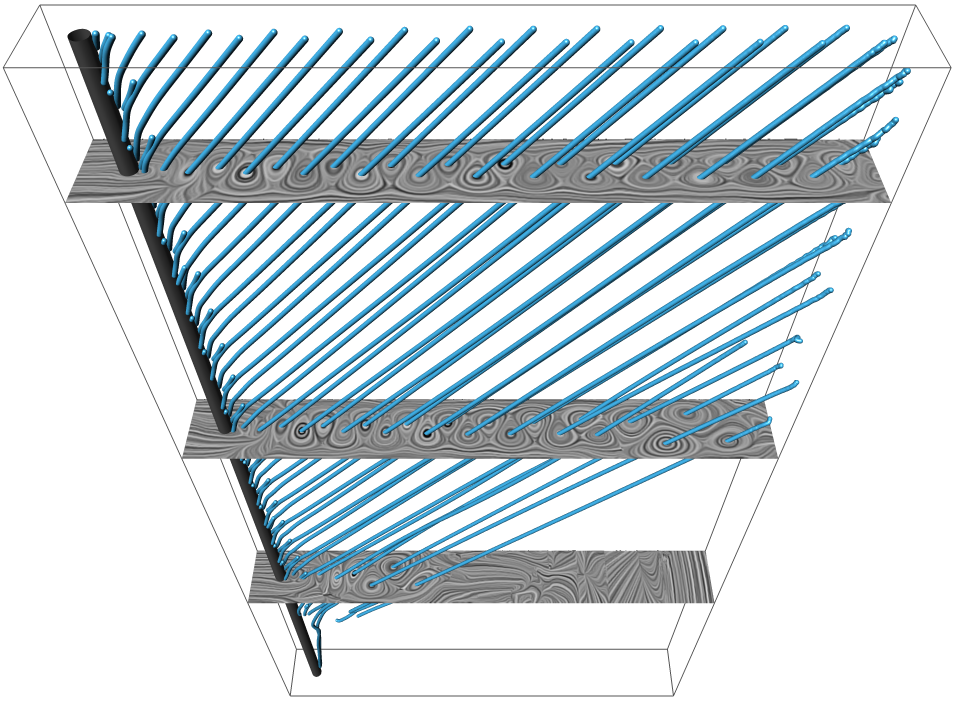}}\vspace{-3mm}%
\caption{Comparison of vortex corelines in space-time in the \dataset{Cylinder} flow.
Despite the noise and resampling artifacts our CNN recovers the reference frame well, so that vortex corelines can be extracted as cleanly as with the linear optimization~\cite{Guenther17:Objective} on the original data.}%
\label{fig:spacetime}%
\end{figure*}

To demonstrate the temporal consistency of our results, we show the paths of vortex cores in space-time in Fig.~\ref{fig:spacetime}.
While the vortex corelines resulting from the linear optimization are disconnected and partially missing on noisy data, our method finely extracts the corelines, which are almost identical to the results of the linear optimization on the original data.

\subsection{Performance}
In the following, we provide timing measurements of our CNN-based approach, compared to the linear optimization~\cite{Guenther17:Objective}.
All measurements were taken with an 8GB NVIDIA GTX 1080 GPU and an Intel i7-6700K CPU at 4.00 GHz with 32 GB memory.
\review{Generating the 30,000 training patches took in total about 37 seconds (24 seconds to compute the unsteady flow patches and 13 seconds to compute the steady ground truth).}
The training takes roughly 10 minutes and the size of the resulting network model is 20.9 MB in HDF5 file format. 
Our method takes $22 ns$ on average for a feed-forward evaluation of a single patch ($5.63 ms$ per batch).
Our method takes a constant evaluation time for a single batch. However, the batch size is limited by the GPU memory, and it makes our computation time linear in the number of local patches over the whole domain. For instance, in the \dataset{Cylinder} data set a slice batch with resolution $5\times80\times640$, takes about $0.893s$, while the linear optimization method takes $0.366s$. Note, however, that our local patch estimation is scalable to the number of available GPUs.

\section{Discussion}\label{sec:discussion}

\paragraph*{Limitations.}
The generalization capacity of a supervised learning approach is always limited to the data that was seen during training.
At present, our network has not seen obstacles or boundary data.
In the future, we plan to expand the training data to also include wall velocity profiles.
As shown in Appendix~\ref{sec:app-boussinesq}, our parametric mixture model is not yet expressive enough to approximate turbulent and small-scale structures accurately.
\review{
The optimal choice of model parameters $m$ also depends on the patch size, which we currently assume to be constant.
In the future, we would like to explore scale-space approaches that look for optimal frames at different patch sizes.
Since the time partial residual can be evaluated, the best patch size could be selected automatically.
}
Further, we concentrated on 2D time-dependent data.
A natural next step is the extension to 3D.

\paragraph*{Choice of Basis.}
\review{
There are several other options to form the parametric vector field model than the approach we described in Section~\ref{sec:deep-mixture-model}.
Standard basis functions would be imaginable, such as monomial, Chebyshev and Fourier basis functions.
While those options have an orthogonal basis, our primitives have a strong prior, since they are derived from models that were fit to experimental data~\cite{Vatistas91}.
In the context of vector field design, radial basis functions have been used to combine primitives~\cite{zhang2006vector}, which gives more intuitive results than adding them up. 
Whether this is closer to observational data or whether it helps a CNN to pick up the flow patterns remains to be explored in future work.
}

\paragraph*{Noise in Linear Method.}
\review{
Our experiments have shown that noise can have a significant impact on the linear method~\cite{Guenther17:Objective}.
Even though, the linear method fits a reference frame transformation to a spatial neighborhood (which has a spatial smoothing effect), there is no smoothing over time.
Further, the second-order accurate finite differences are particularly prone to noise. 
Adding noise of only $1\%$ to $\vv$ in the \dataset{Cylinder} flow already increased the noise level in the time partial $\vv_t$ by $14\%$.
This increase is also dependent on the temporal resolution, since for higher grid resolution the noise in the time domain is more high-frequent.
We found in Fig.~\ref{fig:comparing-cylinder} that smoothing can help to some extent.
Our CNN-based approach removes the noise implicitly.
The global optimization of Hadwiger et al.~\cite{Hadwiger19} also requires the time partial $\vv_t$.
It remains to be tested, whether a high noise level on $\vv_t$ also influences their result.
}

\section{Conclusion}\label{sec:conclusion}
In this paper, we developed a convolutional neural network that extracts the reference frame in which a given unsteady 2D vector field becomes steady.
In such a reference frame, features such as vortices are no longer hidden by ambient motion.
To increase the robustness to noise and resampling artifacts, we trained our network on distorted inputs, which significantly improves the quality over a linear optimization.
In order to generate the required training data, i.e., patches of vector fields, we developed a parametric vector field mixture model that is based on Vatistas' experimentally obtained vortex velocity profile.
To parameterize the model, we fitted it to numerical data and afterwards sampled thousands of training data sets.
Our work shows the great potential of deep learning for end-to-end combinations of preprocessing and feature extraction.

In the future, we would like to increase the generality of our method, for instance by training on other classes of reference frame invariances, such as affine invariance.
We would like to incorporate further training data to obtain more generality, such as for modeling obstacles and boundaries.
Finally, we plan to further improve our parametric mixture model \review{by studying other basis functions and by automatically selecting the patch size in a scale-space approach}.

\paragraph*{Acknowledgements.}
This work was supported by the Swiss National Science Foundation Ambizione grant no.
PZ00P2\_180114.

\appendix
\section{Comparison of Distance Metrics}\label{sec:parameterfitting}
\begin{figure}
\centering
\raisebox{0.12in}{\rotatebox[origin=t]{90}{\tiny $L_2,\lambda=0$}}~%
\includegraphics[width=0.95\columnwidth]{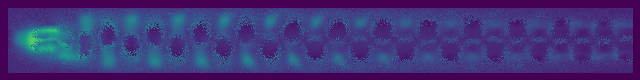}\vspace{0.5mm}\\%
\raisebox{0.12in}{\rotatebox[origin=t]{90}{\tiny$L_1,\lambda=0$}}~%
\includegraphics[width=0.95\columnwidth]{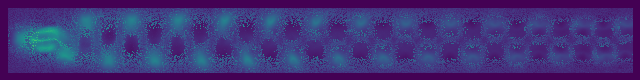}\vspace{0.5mm}\\%
\raisebox{0.12in}{\rotatebox[origin=t]{90}{\tiny$L_1,\lambda=1$}}~%
\includegraphics[width=0.95\columnwidth]{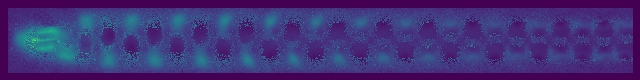}%
\caption{Heat maps of the fitting residual for different vector field distance functions. The mean relative distances are 0.2028, 0.1947 and 0.1906, from top to bottom.}
\label{fig:losses}
\end{figure}

For the fitting of our parametric mixture model to numerical data, we used the distance metric of Kim et al.~\cite{kim18b}, cf. Eq.~\eqref{eq:gradient-loss}.
Following their recommendation, we chose the $L_1$ distance with the gradient weight $\lambda=1$.
Fig.~\ref{fig:losses} presents fitting results for other \review{parameter choices}, showing that the chosen distance metric is best.

\section{Further Fitting Results}
\label{sec:app-boussinesq}
\begin{figure}[t]
\centering
\stackunder{\includegraphics[width=0.32\columnwidth]{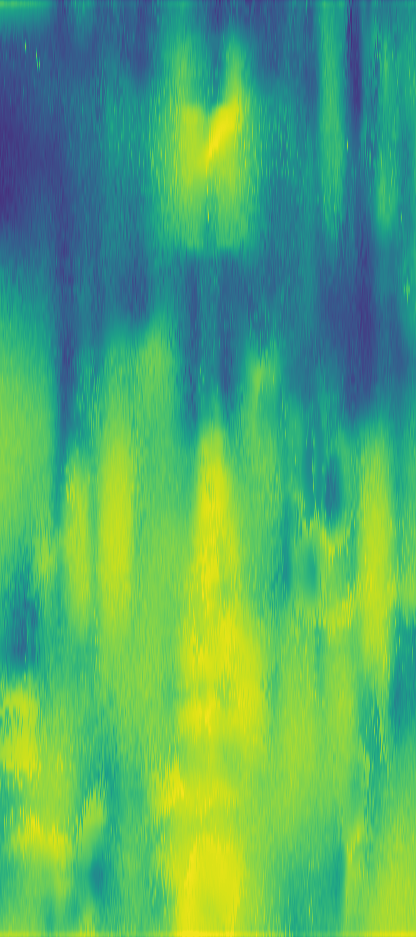}}{{$m=1$}}
\stackunder{\includegraphics[width=0.32\columnwidth]{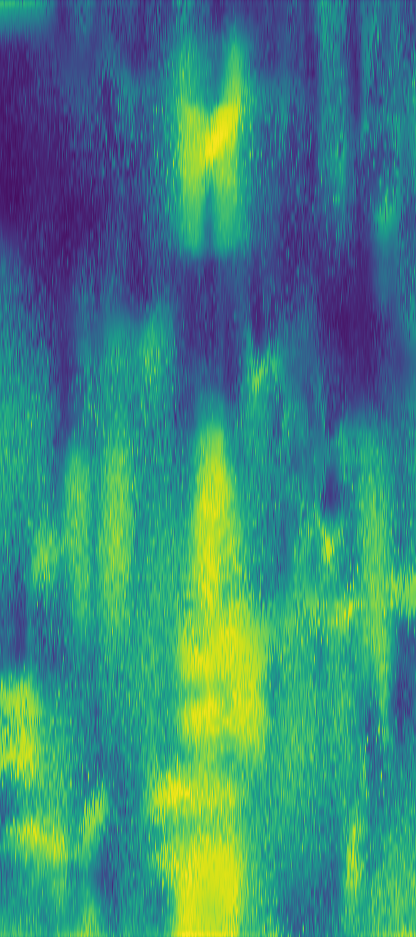}}{{$m=2$}}
\stackunder{\includegraphics[width=0.32\columnwidth]{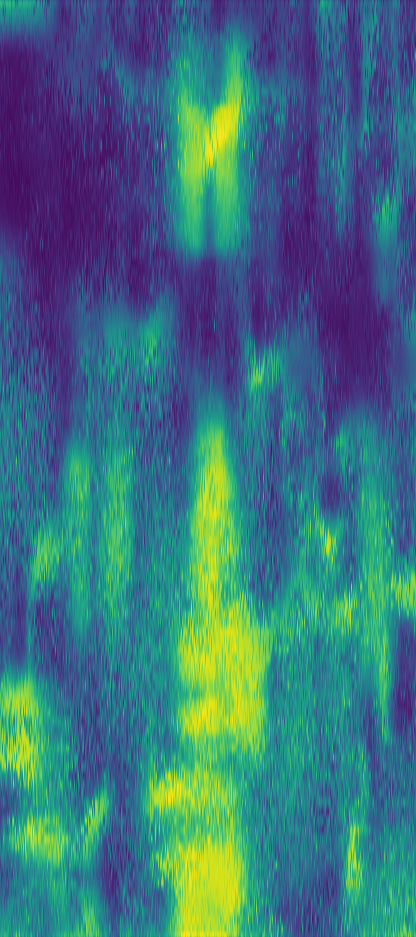}}{{$m=3$}}
\caption{Heat maps of the fitting residual in the \dataset{Boussinesq} data set for varying number of mixture model components $m$.}
\label{fig:fitting-boussinesq}
\end{figure}

In addition to the \dataset{Cylinder} flow, we fitted our parametric mixture model to a more turbulent \dataset{Boussinesq} flow.
This data set contains smaller structures that are more difficult to fit with only $m=3$ mixture model components.
From the results in Fig.~\ref{fig:fitting-boussinesq}, we conclude that the optimal choice of $m$ is data-dependent.

\begin{table}[!b]
\centering
\definecolor{rowblue}{RGB}{220,230,240}
\setlength{\tabcolsep}{5pt} 
\setlength\extrarowheight{1.8pt}
\small
\rowcolors{2}{white}{rowblue}
\resizebox{\linewidth}{!}{	
\begin{tabular}{lll}
\textsc{Parameter} & \textsc{Description} & \textsc{Value}\\ 
\midrule
\midrule
$\Omega$ & Domain & $[-1,1]\times[-1,1]\times[-1,1]$\\
$T,H,W$ & Resolution of the domain $\Omega$ (time, height and width) & $5\times16\times16$\\
$\dot \theta$ & Range of the first-order derivative of the rotation $\dot \mQ$ & $[-0.3,0.3]$\\
$\ddot \theta$ & Range of the second-order derivative of the rotation $\ddot \mQ$ & $[-0.01,0.01]$\\
$\dot x, \dot y$ & Range of the first-order derivative of the translation $\dot \vc$ & $[-0.3,0.3]\times[-0.3,0.3]$\\
$\ddot x, \ddot y$ & Range of the second-order derivative of the translation $\ddot \vc$ & $[-0.01,0.01]\times[-0.01,0.01]$\\
$\alpha$ & Range of uniform noise & $[-0.01,0.01]$\\
$\beta$ & Down-up resampling factor & $0.5$ \\\bottomrule
\end{tabular}%
}%
\vspace{-1mm}%
\normalsize%
\caption{Common parameters for data synthesis and analysis.}
\label{tab:parameters}
\end{table}

\section{Parameters for data synthesis}\label{sec:parameters}

Table~\ref{tab:parameters} lists the parameters of our reference frame transformations. The domain is scaled to the unit box, and the resolution is set to $5\times16\times16$ both for computational efficiency and since objective reference frame transformations are local operations. In fact, the temporal derivative to minimize only depends on up to second-order derivatives~\cite{Guenther17:Objective}. Given the unit domain and its resolution, parameters for the transformation are set at a moderate level.

\bibliographystyle{eg-alpha}

\bibliography{deepvort}

\end{document}